\documentclass[12pt,peerreview]{IEEEtran} 

\usepackage{epsf,cite,amsmath,amscd,amssymb,graphicx,latexsym,multicol}
\begin{document}
\bibliographystyle{ieeetr}

\title{\huge Energy-Efficient Resource Allocation in Wireless Networks
 with Quality-of-Service Constraints\vspace{0.9cm}}

\author{Farhad Meshkati, H. Vincent Poor, Stuart C. Schwartz, and Radu V. Balan\vspace{1cm}
\thanks{This research was supported by the National Science Foundation
under Grant ANI-03-38807. Parts of this work were presented at the
2006 International Wireless Communications and Mobile
Computing Conference (IWCMC), Vancouver, Canada,  July 2006.}
\thanks{F.~Meshkati was with the Department of Electrical Engineering at Princeton University.
He is currently with Qualcomm Inc., 5775 Morehouse Dr., San Diego,
CA 92121 USA (e-mail: meshkati@qualcomm.com).} \thanks{H.~V.~Poor
and S.~C.~Schwartz are with the Department of Electrical
Engineering, Princeton University, Princeton, NJ 08544 USA
(e-mail: {\{poor,stuart\}@princeton.edu}). }
\thanks{R.~V.~Balan is with Siemens Corporate Research, 755 College Road East, Princeton, NJ 08540 USA
(e-mail: radu.balan@siemens.com).}}

\newtheorem{proposition}{Proposition}
\newenvironment{thmproof}[1]
{\noindent\hspace{2em}{\it #1 }}
{\hspace*{\fill}~\QED\par\endtrivlist\unskip}

\centerfigcaptionstrue

\maketitle

\begin{abstract}
A game-theoretic model is proposed to study the cross-layer
problem of joint power and rate control with quality of service
(QoS) constraints in multiple-access networks. In the proposed
game, each user seeks to choose its transmit power and rate in a
distributed manner in order to maximize its own utility while
satisfying its QoS requirements. The user's QoS constraints are
specified in terms of the average source rate and an upper bound
on the average delay where the delay includes both
\emph{transmission} and \emph{queuing} delays. The utility
function considered here measures energy efficiency and is
particularly suitable for wireless networks with energy
constraints. The Nash equilibrium solution for the proposed
non-cooperative game is derived and a closed-form expression for
the utility achieved at equilibrium is obtained. It is shown that
the QoS requirements of a user translate into a ``size" for the
user which is an indication of the amount of network resources
consumed by the user. Using this competitive multiuser framework,
the tradeoffs among throughput, delay, network capacity and energy
efficiency are studied. In addition, analytical expressions are
given for users' delay profiles and the delay performance of the
users at Nash equilibrium is quantified.
\end{abstract}


\section{Introduction}\label{introduction}

Future wireless networks are expected to support a variety of
services with diverse quality of service (QoS) requirements. Because
of the hostile characteristics of wireless channels and scarcity of
radio resources such as power and bandwidth, efficient resource
allocation schemes are necessary for design of high-performance
wireless networks. The objective is to use the radio resources as
efficiently as possible and at the same time satisfy the QoS
requirements of the users in the network. QoS is expressed in terms
of constraints on rate, delay or fidelity. Since in most practical
scenarios, the users' terminals are battery-powered, energy
efficient resource allocation is crucial to prolonging the battery
life of the terminals.

In this work, we study the cross-layer problem of QoS-constrained
joint power and rate control in wireless networks using a
game-theoretic framework. We consider a multiple-access network and
propose a non-cooperative game in which each user seeks to choose
its transmit power and rate in such a way as to maximize its
energy-efficiency (measured in bits per Joule) and at the same time
satisfy its QoS requirements. The QoS constraints are in terms of
the average source rate and the upper bound on the average total
delay (transmission plus queuing delay).  We derive the Nash
equilibrium solution for the proposed game and use this framework to
study trade-offs among throughput, delay, network capacity and
energy efficiency. Network capacity here refers to the maximum
number of users that can be accommodated by the network. While the
delay QoS considered here is in terms of average delay, we also
derive analytical expressions for the user's delay profile  and
quantify the delay performance at Nash equilibrium.

Joint power and rate control with QoS constraints have been studied
extensively for multiple-access networks (see for example
\cite{Honig96} and \cite{Oh99}). In \cite{Honig96}, the authors
study joint power and rate control under bit-error rate (BER) and
average delay constraints. \cite{Oh99} considers the problem of
globally optimizing the transmit power and rate to maximize
throughput of non-real-time users and protect the QoS of real-time
users. Neither work takes into account energy-efficiency. Recently
tradeoffs between energy efficiency and delay have gained more
attention. The tradeoffs in the single-user case are studied in
\cite{Collins99, Prabhakar01, Berry02, Fu03}. The multiuser problem
in turn is considered in \cite{Uysal02} and \cite{Coleman04}. In
\cite{Uysal02}, the authors present a centralized scheduling scheme
to transmit the arriving packets within a specific time interval
such that the total energy consumed is minimized whereas in
\cite{Coleman04}, a distributed ALOHA-type scheme is proposed for
achieving energy-delay tradeoffs.  Joint power and rate control for
maximizing goodput in delay-constrained networks is studied in
\cite{Ahmed04}.

Recently, game theory has been used for studying power control in
code-division-multiple-access (CDMA) networks \cite{JiHuang98,
Shah98, GoodmanMandayam00, Xiao01, Zhou01, Alpcan, Feng01, Sung,
Saraydar01, Saraydar02, Gunturi03, AA03, MeshkatiTcomm,
MeshkatiJSAC, MeshkatiISIT05}). Each user seeks to choose its
transmit power in order to maximize its utility. In \cite{Alpcan}
and \cite{Gunturi03}, the utility function in \eqref{eq5} is
chosen for the users and the corresponding Nash equilibrium
solution is derived. In \cite{Shah98} and
\cite{GoodmanMandayam00}, the authors use a utility function that
measures the number of reliable bits that are transmitted per
joule of energy consumed. The analysis is extended in
\cite{Saraydar02} by introducing pricing to improve the efficiency
of Nash equilibrium. Joint energy-efficient power control and
receiver design is studied in \cite{MeshkatiTcomm}. In addition, a
game-theoretic approach to energy-efficient power allocation in
multicarrier systems is presented in \cite{MeshkatiJSAC}. Joint
network-centric and user-centric power control is discussed in
\cite{Feng01}. In \cite{Sung}, the utility function is assumed to
be proportional to the user's throughput and a pricing function
based on the normalized received power of the user is proposed.
S-modular power control games are studied in \cite{AA03}. The
prior work in this area does not explicitly take into account the
QoS requirements of the users. While \cite{MeshkatiISIT05}
proposes a delay-constrained power control game,  it considers the
transmission delay only and does not perform any rate control.

This work is the first study of QoS-constrained power and rate
control in multiple-access networks using a game-theoretic
framework. In our proposed game-theoretic model, users choose
their transmit powers and rates in a \emph{competitive} and
\emph{distributed} manner in order to maximize their energy
efficiency and at the same time satisfy their delay and rate QoS
requirements. Using this framework, we also analyze the tradeoffs
among throughput, delay, network capacity and energy efficiency.
While centralized resource allocation schemes can achieve a better
performance compared to distributed algorithms, in most practical
scenarios, distributed algorithms are preferred over centralized
ones. Centralized algorithms tend to be complex and not easily
scalable. Hence, throughout this article, we focus on distributed
algorithms with emphasis on energy efficiency.

The remainder of this paper is organized as follows. In
Section~\ref{system model}, we describe the system model. The
proposed joint power and rate control game is discussed in
Section~\ref{PRCG} and its Nash equilibrium solution is derived in
Section~\ref{NE}. We then describe an admission control scheme in
Section~\ref{admission control}. The users' delay performance is
analyzed in Section~\ref{delay performance}. Based on our
analysis, the tradeoffs among throughput, delay, network capacity
and energy efficiency are studied in Section~\ref{numerical
results} using numerical results. Finally, we give conclusions in
Section~\ref{conclusions}.

\section{System Model}\label{system model}

We consider a direct-sequence CDMA (DS-CDMA) network and propose a
non-cooperative (distributed) game in which each user seeks to
choose its transmit power and rate to maximize its energy
efficiency (measured in bits per joule) while satisfying its QoS
requirements. We specify the QoS constraints of user $k$ by
$(r_k,D_k)$ where $r_k$ is the average source rate and $D_k$ is
the upper bound on average delay. The delay includes both queuing
and transmission delays. The incoming traffic is assumed to have a
Poisson distribution with parameter $\lambda_k$ which represents
the average packet arrival rate with each packet consisting of $M$
bits. The source rate (in bit per second), $r_k$, is hence given
by
\begin{equation}\label{eq0}
    r_k= M \lambda_k .
\end{equation}
The user transmits the arriving packets at a rate $R_k$ (bps) and
with a transmit power equal to $p_k$ Watts. We consider an
automatic-repeat-request (ARQ) mechanism in which the user keeps
retransmitting a packet until the packet is received at the access
point without any errors. The incoming packets are assumed to be
stored in a queue and transmitted in a first-in-first-out (FIFO)
fashion. The packet transmission time for user $k$ is defined as
\begin{equation}\label{eq1}
    \tau_k = \frac{M}{R_k} + \epsilon_k \simeq \frac{M}{R_k} ,
\end{equation}
where $\epsilon_k$ represents the time taken for the user to
receive an ACK/NACK from the access point. We assume $\epsilon_k$
is negligible compared to $\frac{M}{R_k}$. The packet success
probability (per transmission) is represented by $f(\gamma_k)$
where $\gamma_k$ is the received signal-to-interference-plus-noise
ratio (SIR) for user~$k$. The retransmissions are assumed to be
independent. The packet success rate, $f(\gamma)$, is assumed to
be increasing and S-shaped\footnote{An increasing function is
S-shaped if there is a point above which the function is concave,
and below which the function is convex.} (sigmoidal) with $f(0)=0$
and $f(\infty)=1$. This is a valid assumption for many practical
scenarios as long as the packet size is reasonably large (e.g.,
$M=100$ bits) \cite{MeshkatiTcomm}.

\begin{figure}
\centering
\includegraphics[width=4in]{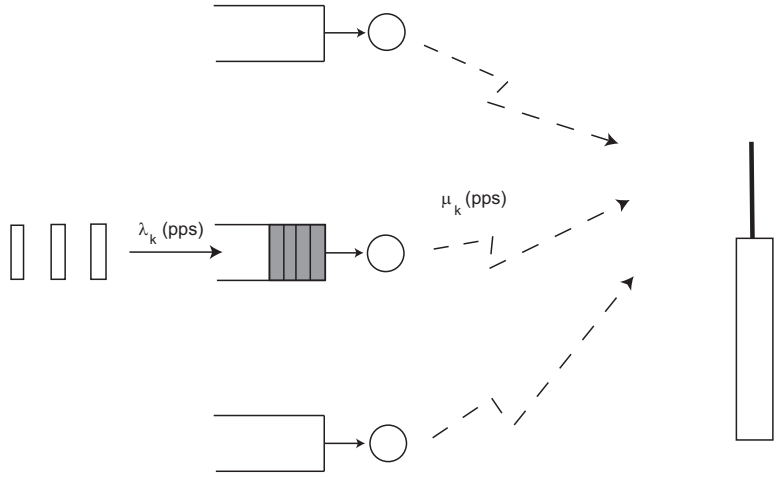}
\caption{System model based on an M/G/1 queue.} \label{fig1-sys}
\end{figure}

We can represent the combination of user $k$'s queue and wireless
link as an M/G/1 queue, as shown in Fig.~\ref{fig1-sys} where the
traffic is Poisson with parameter $\lambda_k$ (in packets per
second) and the service time, $S_k$, has the following probability
mass function (PMF):
\begin{equation}\label{eq2}
    \textrm{Pr}\{S_k=m\tau_k\}= f(\gamma_k)
    \left(1-f(\gamma_k)\right)^{m-1}  \ \ \ \textrm{for} \ m=1, 2,
    \cdots
\end{equation}
As a result, we have
\begin{equation}\label{eq3}
    \mathbb{E}\{S_k\}=\sum_{m=1}^{\infty} m \tau_k
    \left(1-f(\gamma_k)\right)^{m-1} = \frac{\tau_k}{f(\gamma_k)}
    .
\end{equation}
Consequently, the service rate, $\mu_k$, is given by
\begin{equation}\label{eq4}
    \mu_k=\frac{1}{\mathbb{E}\{S_k\}}= \frac{f(\gamma_k)}{\tau_k}
    ,
\end{equation}
and the load factor
$\rho_k=\frac{\lambda_k}{\mu_k}=\frac{\lambda_k
\tau_k}{f(\gamma_k)}$.

To keep the queue of user $k$ stable, we must have $\rho_k<1$ or
$f(\gamma_k)>\lambda_k \tau_k$. Now, let $W_k$ be a random
variable representing the total packet delay for user $k$. This
delay includes the time the packet spends in the queue,
$W^{(q)}_k$, as well as the service time, $S_k$. Hence, we have
\begin{equation}\label{eq4b}
    W_k = W^{(q)}_k + S_k .
\end{equation}
It is known that for an M/G/1 queue the average wait time (including
the queuing and service time) is given by
\begin{equation}\label{eq5}
    \bar{W}_k=\frac{\bar{L}_k}{\lambda_k} ,
\end{equation}
where $\bar{L}_k= \rho_k + \frac{ \rho_k^2 + \lambda_k^2
\sigma_{S_k}^2}{2(1-\rho_k)}$ with $\sigma_{S_k}^2$ being the
variance of the service time \cite{GrossBook85}. Therefore, the
average packet delay for user $k$ is given by
\begin{equation}\label{eq6}
    \bar{W}_k = \tau_k \left(\frac{1-\frac{\lambda_k
    \tau_k}{2}}{f(\gamma_k)-\lambda_k \tau_k}\right) \ \ \
    \textrm{with} \ f(\gamma_k)>\lambda_k \tau_k .
\end{equation}
We require the average packet delay for user $k$ to be less than
or equal to $D_k$, i.e.,
\begin{equation}\label{eq6b}
    \bar{W}_k \leq D_k
\end{equation}
This translates to
\begin{equation}\label{eq7}
    f(\gamma_k) \geq \lambda_k \tau_k + \frac{\tau_k}{D_k} -
    \frac{\lambda_k \tau_k^2}{2D_k} .
\end{equation}
However, since $0\leq f(\gamma_k) \leq 1$, we must
have\footnote{Note that $f(\gamma)=1$ requires an infinite SIR
which is not practical.}
\begin{equation}\label{eq8}
  0 \leq \lambda_k \tau_k + \frac{\tau_k}{D_k} -
    \frac{\lambda_k \tau_k^2}{2D_k} < 1.
\end{equation}
This means that $r_k=M\lambda_k$ and $D_k$ are feasible if only if
they satisfy \eqref{eq8}. Note that since the upper bound on the
average delay cannot be smaller than the transmission time, i.e.,
$\frac{D_k}{\tau_k}\geq1$, then we must have $R_k\geq M/D_k$. This
automatically implies that $\lambda_k \tau_k + \frac{\tau_k}{D_k}
-\frac{\lambda_k \tau_k^2}{2D_k}>0$.

Let us define $\eta_k=\lambda_k \tau_k + \frac{\tau_k}{D_k}
-\frac{\lambda_k \tau_k^2}{2D_k}$. Then, \eqref{eq7} is equivalent
to the condition $\gamma\geq \hat{\gamma}_k$ where
\begin{equation}\label{eq9}
    \hat{\gamma}_k=f^{-1}(\eta_k) \ ,
\end{equation}
with $\eta_k<1$ and $R_k\geq M/D_k$. This means that the delay
constraint in \eqref{eq6b} translated into a lower bound on the
output SIR.

\section{The Joint Power and Rate Control Game} \label{PRCG}

Consider the non-cooperative joint power and rate control game
(PRCG) $\mathcal{G}=[\mathcal{K}, \{\mathcal{A}_k\}, \{u_k\}]$
where $\mathcal{K}=\{1,2,\cdots,K\}$ is the set of users,
$\mathcal{A}_k=[0,P_{max}]\times[0,B]$ is the strategy set for
user $k$ with a strategy corresponding to a choice of transmit
power and transmit rate, and $u_k$ is the utility function for
user $k$. Here, $P_{max}$ and $B$ are the maximum transmit power
and the system bandwidth, respectively. For the sake simplicity,
throughout this paper, we assume $P_{max}$ is large. Each user
chooses its transmit power and rate in order to maximize its own
utility while satisfying its QoS requirements. The utility
function for a user is defined as the ratio of the user's goodput
to its transmit power, i.e.,
\begin{equation}\label{eq10}
    u_k=\frac{T_k}{p_k} ,
\end{equation}
where the goodput $T_k$ is the number of bits that is transmitted
successfully per second and is given by
\begin{equation}\label{eq11}
    T_k= R_k f(\gamma_k) .
\end{equation}
Therefore, the utility function for user $k$ is given by
\begin{equation}\label{eq12}
    u_k=R_k \frac{f(\gamma_k)}{p_k} .
\end{equation}
This utility function, which was first introduced in \cite{Shah98,
GoodmanMandayam00}, has units of bits per joule and is
particularly suitable for wireless networks where energy
efficiency is important.

Fixing the other users' transmit powers and rates, the
utility-maximizing strategy for user $k$ is given by the solution of
the following constrained maximization:
\begin{equation}\label{eq12b}
    \max_{p_k, R_k} \ u_k \ \ \ \textrm{s.t.} \ \ \ \bar{W}_k \leq D_k \ ,
\end{equation}
or equivalently
\begin{equation}\label{eq13}
   \max_{p_k, R_k} \ u_k \ \ \ \ \textrm{s.t.} \ \
   \gamma_k \geq \hat{\gamma}_k
\end{equation}
with $0\leq \eta_k<1$ where
\begin{equation}\label{eq13a}
\hat{\gamma}_k=f^{-1}(\eta_k) ,
\end{equation}
and
\begin{equation}\label{eq13b}
    \eta_k=\frac{r_k}{R_k} + \frac{M}{D_k R_k} -\frac{M r_k}{2D_k
    R_k^2} \ .
\end{equation}
Note that for a matched filter receiver and with random spreading
sequences, the received SIR is approximately given by
\begin{equation}\label{eq14}
    \gamma_k= \left(\frac{B}{R_k}\right) \frac{p_k h_k}{\sigma^2 + \sum_{j\neq
    k} p_j h_j} ,
\end{equation}
where $h_k$ is the channel gain for user $k$ and $\sigma^2$ is the
noise power in the bandwidth $B$.

Let us first look at the maximization in \eqref{eq13} without any
constraints. Based on \eqref{eq14}, we can write
\begin{equation}\label{eq15}
    \max_{p_k, R_k} u_k \ \ \equiv \ \max_{\gamma_k, R_k}  B \hat{h}_k \frac{f(\gamma_k)}{\gamma_k}  .
\end{equation}


\begin{proposition}
The unconstrained utility maximization in \eqref{eq15} has an
infinite number of solutions. More specifically, any combination
of $p_k$ and $R_k$ that achieves an output SIR equal to
$\gamma^*$, the solution to $f(\gamma)=\gamma f'(\gamma)$,
maximizes $u_k$.
\end{proposition}

\begin{thmproof}{Proof:}
Notice from \eqref{eq15} that when the other users' powers and
rates are fixed (i.e., fixed $\hat{h}_k$), user $k$'s utility
depends only on $\gamma$ and is independent of the specific values
of $p_k$ and $R_k$. In addition, by taking the derivative of
$\frac{f(\gamma)}{\gamma}$ with respect to $\gamma$ and equating
it to zero, it can be shown that $\frac{f(\gamma)}{\gamma}$  is
maximized when $\gamma=\gamma^*$, the (unique) positive solution
of $f(\gamma)=\gamma f'(\gamma)$. Therefore, $u_k$ is maximized
for any combination of $p_k$ and $R_k$ for which
$\gamma_k=\gamma^*$. This means that there are infinitely many
solutions for the unconstrained maximization in \eqref{eq15}.
\end{thmproof}

%


Now, considering that $\tau_k= M/R_k$ must be less than or equal
to $D_k$, the condition $0\leq \eta_k<1$ is equivalent to
\begin{equation}\label{eq18}
    R_k > \left(\frac{M}{D_k}\right) \frac{1+D_k \lambda_k
    +\sqrt{1+ D_k^2 \lambda_k^2}}{2} .
\end{equation}

Let us define $$\Omega_k^{\infty}=\left(\frac{M}{D_k}\right)
\frac{1+D_k\lambda_k +\sqrt{1+ D_k^2 \lambda_k^2}}{2}.$$ Note that
for $R_k=\Omega_k^{\infty}$, we have $\eta_k=1$ and hence
$\hat{\gamma}_k=\infty$. Also, define $\Omega_k^*$ as the rate for
which $\hat{\gamma}_k=\gamma^*$, i.e.,
\begin{equation} \label{eq20}
\Omega_k^*=\left(\frac{M}{D_k}\right) \frac{1+D_k\lambda_k +\sqrt{1+
D_k^2 \lambda_k^2 +2(1-f^*)D_k \lambda_k}}{2f^*}
\end{equation}
where $f^*=f(\gamma^*)$. It is straightforward to show that
$\hat{\gamma}_k$ is a decreasing function of $R_k$ for all
$R_k\geq\Omega_k^{\infty}$. Therefore, $\hat{\gamma}_k
> \gamma^*$ for all $\Omega_k^{\infty}\leq R_k <\Omega_k^*$.  This
means that user $k$ has no incentive to transmit at a rate smaller
than $\Omega_k^*$. Furthermore, based on Proposition~1, any
combination of $p_k$ and $R_k \geq \Omega_k^*$ which results in an
output SIR equal to $\gamma^*$ is a solution to the constrained
maximization in \eqref{eq13}. Note that when $R_k = \Omega_k^*$
and $\gamma_k=\gamma^*$, we have $\bar{W}_k=D_k$.

If $\gamma^*$ is not feasible due to the maximum transmit power
limitation, the user has to adjust its transmission rate and
target SIR to satisfy its QoS constraints. In particular, user $k$
would choose $\tilde{\Omega}_k$ as its transmission rate such that
its transmit rate and target SIR such that
$$\tilde{\Omega}_k=\left(\frac{M}{D_k}\right) \frac{1+D_k\lambda_k
+\sqrt{1+ D_k^2 \lambda_k^2 +2\left(1-f(\tilde{\gamma})\right) D_k
\lambda_k}}{2 f(\tilde{\gamma}_k)}$$ where
$$\tilde{\gamma}_k = (B/\tilde{\Omega}_k) P_{max} \hat{h}_k.$$
This, of course, results in a reduction in the user's energy
efficiency.

\section{Nash Equilibrium for the PRCG} \label{NE}

For a non-cooperative game, a Nash equilibrium is defined as a set
of strategies for which no user can unilaterally improve its own
utility \cite{FudenbergTiroleBook91}. We saw in Section~\ref{PRCG}
that for our proposed non-cooperative game, each user has infinitely
many strategies that maximize the user's utility. In particular, any
combination of $p_k$ and $R_k$ for which $\gamma_k=\gamma^*$ and
$R_k\geq\Omega_k^*$ is a best-response strategy.

\begin{proposition}
If  $\sum_{k=1}^K \frac{1}{1+\frac{B}{\Omega_k^* \gamma^*}} < 1$,
then the PRCG has at least one Nash equilibrium given by $(p_k^*,
\Omega_k^*)$, for $k=1,\cdots,K$, where
$p_k^*=\frac{\sigma^2}{h_k}\left(\frac{\frac{1}{1+\frac{B}{\Omega_k^*\gamma^*}}}{1-\sum_{j=1}^K
\frac{1}{1+\frac{B}{\Omega_j^*\gamma^*}}}\right)$ and $\Omega_k^*$
is given by \eqref{eq20}. Furthermore, when there are more than
one Nash equilibrium, $(p_k^*, \Omega_k^*)$ is the Pareto-dominant
equilibrium.
\end{proposition}

\begin{thmproof}{Proof:}
If $\sum_{j=1}^K \frac{1}{1+\frac{B}{\Omega_j^* \gamma^*}} < 1$
then
$p_k^*=\frac{\sigma^2}{h_k}\left(\frac{\frac{1}{1+\frac{B}{\Omega_k^*\gamma^*}}}{1-\sum_{j=1}^K
\frac{1}{1+\frac{B}{\Omega_j^*\gamma^*}}}\right)$ is positive and
finite. Now, if we let $p_k=p_k^*$ and $R_k=\Omega_k^*$, then the
output SIR for all the users will be equal to $\gamma^*$ which
means every user is using its best-response strategy. Therefore,
$(p_k^*, R_k^*)$ for $k=1,\cdots,K$ is a Nash equilibrium.

More generally, if we let $R_k= \tilde{R}_k \geq \Omega_k^*$ and
provided that $\sum_{j=1}^K \frac{1}{1+\frac{B}{\tilde{R}_j
\gamma^*}} < 1$, then $(\tilde{p}_k, \tilde{R}_k)$ is a Nash
equilibrium where
$\tilde{p}_k=\frac{\sigma^2}{h_k}\left(\frac{\frac{1}{1+\frac{B}{\tilde{R}_k
\gamma^*}}}{1-\sum_{j=1}^K \frac{1}{1+\frac{B}{\tilde{R}_j
\gamma^*}}}\right)$.

Based on \eqref{eq12}, at Nash equilibrium, the utility of user $k$
is given by
\begin{eqnarray}\label{eq20b}
    u_k&=& \frac{B f(\gamma^*) h_k}{\sigma^2
\gamma^*}\left(\frac{1-\sum_{j=1}^K \frac{1}{1+\frac{B}{\tilde{R}_j
\gamma^*}}}{1-\frac{1}{1+\frac{B}{\tilde{R}_k \gamma^*}}}\right)
\nonumber\\ &=& \frac{B f(\gamma^*) h_k}{\sigma^2 \gamma^*}\left(1-
\frac{\sum_{j \neq k} \frac{1}{1+\frac{B}{\tilde{R}_j
\gamma^*}}}{1-\frac{1}{1+\frac{B}{\tilde{R}_k \gamma^*}}}\right) \ .
\end{eqnarray}

Therefore, the Nash equilibrium with the smallest $\tilde{R}_k$
achieves the largest utility. A higher transmission rate for a
user requires a larger transmit power by that user to achieve
$\gamma^*$. This not only reduces the user's utility but also
causes more interference for other users in the network and forces
them to raise their transmit powers as well which will result in a
reduction in their utilities. This means that the Nash equilibrium
with $R_k=\Omega_k^*$ and $p_k^*$ for $k=1,\cdots,K$ is the
Pareto-efficient Nash equilibrium.
\end{thmproof}


We define the ``size" of user $k$ as
\begin{equation}\label{eq24}
    \Phi_k^* = \frac{1}{1+\frac{B}{\Omega_k^* \gamma^*}} \ .
\end{equation}
Based on this definition, the feasibility condition $\sum_{k=1}^K
\frac{1}{1+\frac{B}{\Omega_k^* \gamma^*}} < 1$ can be written as
\begin{equation}\label{eq25}
    \sum_{k=1}^K \Phi_k^* < 1 .
\end{equation}
%
%

Note that the QoS requirements of user $k$ (i.e., its source rate
$r_k$ and delay constraint $D_k$) uniquely determine $\Omega_k^*$
through \eqref{eq20} and, in turn, determine the size of the user
(i.e., $\Phi_k^*$) through \eqref{eq24}. The size of a user is
basically an indication of the amount of network resources consumed
by that user. A larger source rate or a tighter delay constraint for
a user increases the size of the user. The network can accommodate a
set of users if and only if their total size is less than 1. In
Section~\ref{numerical results}, we use this framework to study the
tradeoffs among throughput, delay, network capacity and energy
efficiency.

\section{Admission Control}\label{admission control}


In Section~\ref{NE}, we defined the ``size" of a user based on its
QoS requirements. Before joining the network, each user calculates
its size using \eqref{eq24} and announces it to the access point.
According to \eqref{eq25}, the access point admits those users whose
total size is less than 1. While the goal of each user is to
maximize its own energy efficiency, a more sophisticated admission
control can be performed to maximize the total network utility. In
other words, out of the $K$ users, the access point can choose those
users for which the total network utility is the largest, i.e.,
\begin{equation}\label{eq26}
    \max_{{\mathcal{L}}\subset \{1,\cdots,K\}} \ \sum_{\ell \in {\mathcal{L}}} u_{\ell}
\end{equation}
under the constraint that $\sum_{\ell \in {\mathcal{L}}}
\Phi_{\ell}^* <1$.

Based on \eqref{eq20b}, the utility of user $\ell$ at the
Pareto-dominant Nash equilibrium is given by
\begin{equation}\label{eq28}
u_{\ell}=\left(\frac{B h_{\ell} f(\gamma^*)}{\sigma^2
\gamma^*}\right)\frac{1-\sum_{i \in
    {\mathcal{L}}}\Phi_i^*}{1-\Phi_{\ell}^*} \ .
\end{equation}
As a result, \eqref{eq26} becomes
$$\max_{{\mathcal{L}}\subset \{1,\cdots,K\}} \ \sum_{\ell \in
     {\mathcal{L}}} h_{\ell} \frac{1-\sum_{i \in
    {\mathcal{L}}}\Phi_i^*}{1-\Phi_{\ell}^*}$$ or equivalently
\begin{equation}\label{eq29}
      \max_{{\mathcal{L}}\subset \{1,\cdots,K\}} \left(1-\sum_{i \in
    {\mathcal{L}}}\Phi_i^*\right)  \sum_{\ell \in
     {\mathcal{L}}}  \frac{h_{\ell}}{1-\Phi_{\ell}^*}\
\end{equation}
under the constraint that $\sum_{\ell \in {\mathcal{L}}}
\Phi_{\ell}^* <1$.

In general, obtaining a closed-form solution for \eqref{eq29} is
difficult. Instead, in order to gain some insight, let us consider
the special case in which all users are at the same distance from
the access point. We first consider the scenario in which the
users have identical QoS requirements (i.e.,
$\Phi_1^*=\cdots=\Phi_K^*=\Phi^*$). If we replace $\sum_{\ell=1}^L
h_\ell$ by $L \mathbb{E}\{h\}$, then \eqref{eq29} becomes
\begin{equation}\label{eq30}
    \max_L \frac{\mathbb{E}\{h\} ( L-L^2\Phi^* )}{1-\Phi^*} .
\end{equation}
Therefore, the optimal number of users for maximizing the total
utility in the network is $L=\left[\frac{1}{2\Phi^*}\right]$ where
$[x]$ represents the integer nearest to $x$.

Now consider another scenario in which there are $C$ classes of
users. The users in class $c$ are assumed to all have the same QoS
requirements and hence the same size, $\Phi^{*(c)}$. Since we are
assuming that all the users have the same distance from the access
point, they all have the same channel gains. Now, if the access
point admits $L^{(c)}$ users from class $c$ then the total utility
is given by
$$u_{T}=\left(\frac{B h f(\gamma^*)}{\sigma^2 \gamma^*}\right)
\left(1-\sum_{c=1}^C L^{(c)}\Phi^{*(c)}\right)\left(\sum_{c=1}^C
\frac{L^{(c)}}{1-\Phi^{*(c)}}\right)$$ provided that $\sum_{c=1}^C
L^{(c)}\Phi^{*(c)}<1$. Without loss of generality, let us assume
that $\Phi^{*(1)}<\Phi^{*(2)}< \cdots <\Phi^{*(C)}$. It can be
shown that $u_T$ is maximized when
$L^{(1)}=\left[\frac{1}{2\Phi^{*(1)}}\right]$ with $L^{(c)}=0$ for
$c=2, 3, \cdots, C$. This is because adding a user from class~1 is
always more beneficial in terms of increasing the total utility
than adding a user from any other class. Therefore, in order to
maximize the total utility in the network, the access point should
admit only users from the class with the smallest size. While this
solution maximizes the total network utility, it is not fair. A
more sophisticated admission control mechanism can be used to
improve the fairness.

\section{Delay Performance}\label{delay performance}

In Section~\ref{system model}, we defined the delay requirement of
a user as an upper bound on the average total packet delay for
that user where the total delay, $W_k$, is given by the sum of the
queuing delay and service time. We have considered a scenario in
which users choose their transmit powers and rates in a selfish
and distributed manner such that they maximize their own energy
efficiency while satisfying their delay requirements. In
Section~\ref{NE}, we showed that at the Pareto-dominant Nash
equilibrium, the transmit power and rate of a user are such that
the delay bound is met with equality. However, it would be useful
to obtain the delay profile of a user so that the deviations of
the true delay from the average value can be quantified. More
specifically, we would like to find a closed-form expression for
$\textrm{Pr}\{W_k \leq c \}$ for all $c$.

To that end, let us define $w_k(t)$ as the probability density
function (PDF) of $W_k$. Then, we have
\begin{equation}\label{eq31a}
\textrm{Pr}\{W_k \leq c \}= \int_0^c w_k(t) \textrm{d}t \ .
\end{equation}
Let $W_k^*(s)$ represent the Laplace transform for $w_k(t)$, i.e.,
\begin{equation}\label{eq31}
    W_k^*(s)= \int_0^\infty e^{-st} w_k(t) \textrm{d}t \ .
\end{equation}
It is known that for M/G/1 queues, we have
\begin{equation}\label{eq32}
    W_k^*(s)=\frac{(1-\rho_k) s B_k^*(s)}{s- \lambda_k [1-B_k^*(s)]}
\end{equation}
where $B_k^*(s)=\int_0^\infty e^{-st} b_k(t) \textrm{d}t$ with
$b_k(t)$ being the PDF of the service time $S_k$ \cite{GrossBook85}.
Based on \eqref{eq2}, $b_k(t)$ is given by
\begin{equation}\label{eq33}
    b_k(t)=\sum_{m=1}^\infty f(\gamma_k)
    \left(1-f(\gamma_k)\right)^{m-1} \delta (t-m\tau_k)
\end{equation}
where $\delta(\cdot)$ is the Dirac delta function. Therefore, we
have
\begin{equation}\label{eq34}
    B_k^*(s)=\frac{f(\gamma_k)}{e^{s\tau_k} -1+f(\gamma_k)} .
\end{equation}
As a result,
\begin{equation}\label{eq34b}
    W_k^*(s)=\frac{(1-\rho_k)f(\gamma_k)s}{s\left(e^{s\tau_k} -1 +
    f(\gamma_k)\right)-\lambda_k \left(e^{s\tau_k} -1\right)} .
\end{equation}
However, obtaining a closed-form expression for $w_k(t)$ based on
$W_k^*(s)$ in \eqref{eq34b} is very difficult. But, recall from
Section~\ref{system model} that $$W_k=W_k^{(q)}+S_k.$$ Based on this
we have
\begin{equation}\label{eq34c}
W_k^{(q)*}(s)=\frac{W_k^*(s)}{B_k^*(s)}=
\frac{(1-\rho_k)s\left(e^{s\tau_k} -1 +
    f(\gamma_k)\right)}{s\left(e^{s\tau_k} -1 +
    f(\gamma_k)\right)-\lambda_k \left(e^{s\tau_k} -1\right)} .
\end{equation}
While finding the inverse Laplace transform of \eqref{eq34c} is also
difficult, we will shortly derive an accurate approximation for
$w_k^{(q)}(t)$. Before doing that, let us first obtain the mean and
variance of $W_k^{(q)}$ and $S_k$. For simplicity of notation, we
will drop the subscript $k$ but it should be noted that all of our
results are user dependent. Also, we replace $f(\gamma)$ by $f$.

\begin{figure}
\centering
\includegraphics[width=5in]{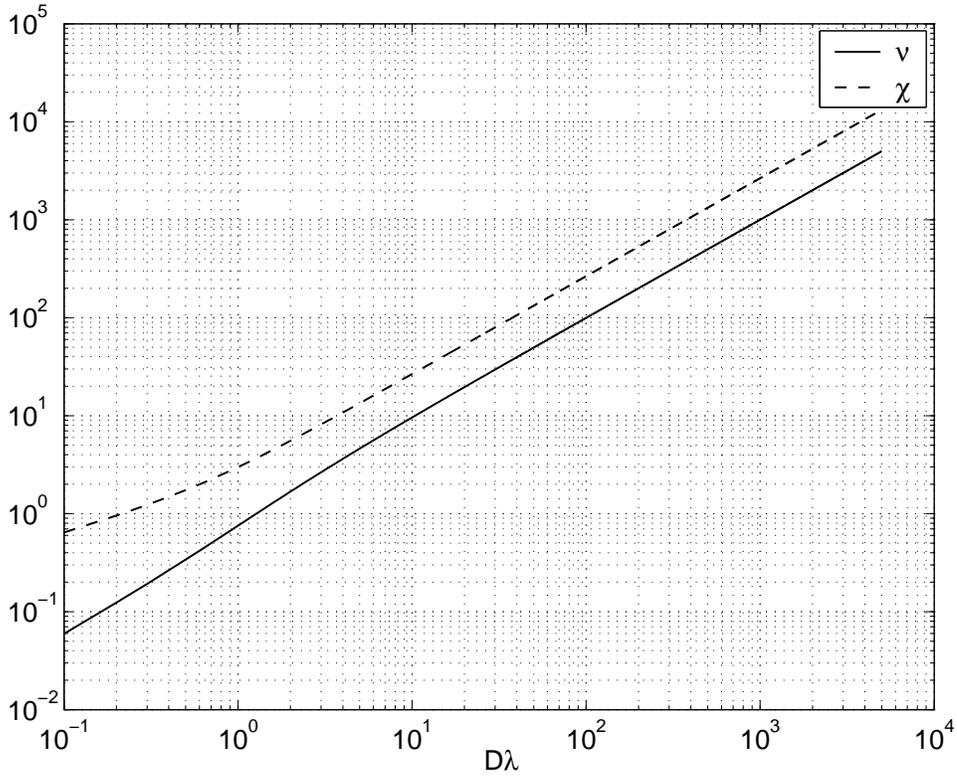}
\caption{Plots of $\nu=\frac{\bar{W}^{(q)}}{\bar{S}}$ and
$\chi=\sqrt{\frac{\sigma_{W^{(q)}}^2}{\sigma_S^2}}$ as a function
of $d=D\lambda$. $\lambda$ is the average source rate in packets
per second and $D$ is the average delay bound in seconds.}
\label{ratios}
\end{figure}

Based on \eqref{eq2}, the mean and variance of $S$ are,
respectively, given by
\begin{equation}\label{eq34d}
    \bar{S}=\frac{\tau}{f}
\end{equation}
and
\begin{equation}\label{eq34e}
    \sigma_S^2=\frac{\tau^2}{f^2}(1-f).
\end{equation}
From the known properties of M/G/1 queues \cite{GrossBook85}, the
mean and variance of $W^{(q)}$ are, respectively, given by
\begin{equation}\label{eq34f}
    \bar{W}^{(q)}=
\frac{\tau}{f}\left[\frac{(1-\frac{f}{2})\left(\frac{\lambda\tau}{f}\right)}{1-\frac{\lambda\tau}{f}}\right]
\end{equation}
and $$\sigma_{W^{(q)}}^2 = \mathbb{E}\left\{W^{(q)^2}\right\} -
\bar{W}^{(q)^2} = \frac{\lambda}{1-\rho} \left[ \bar{W}^{(q)}
\mathbb{E}\{S^2\} + \frac{\mathbb{E}\{S^3\}}{3} \right] -
\bar{W}^{(q)^2}.$$ After some manipulations, it can be shown that
the variance of $W^{(q)}$ is given by
\begin{equation}\label{eq34g}
    \sigma_{W^{(q)}}^2
=\frac{\tau^2}{f^2}(1-f) \left[
\frac{1}{\left(1-\frac{\lambda\tau}{f}\right)^2} + \frac{f^2
\left(\frac{\lambda\tau}{f}\right)
\left(4-\frac{\lambda\tau}{f}\right)}{12(1-f)(1-\frac{\lambda\tau}{f})^2}
-1\right] .
\end{equation}
To gain some insights into the contributions of the queuing delay
and service time to the overall delay, let us define
$$\nu=\frac{\bar{W}^{(q)}}{\bar{S}}$$ and
$$\chi=\sqrt{\frac{\sigma_{W^{(q)}}^2}{\sigma_S^2}}.$$ Then, we have
\begin{equation}\label{eq35}
    \nu=\frac{(1-\frac{f}{2})\left(\frac{\lambda\tau}{f}\right)}{1-\frac{\lambda\tau}{f}}
    \ ,
\end{equation}
and
\begin{equation}\label{eq36}
    \chi=\left[
\frac{1}{\left(1-\frac{\lambda\tau}{f}\right)^2} + \frac{f^2
\left(\frac{\lambda\tau}{f}\right)
\left(4-\frac{\lambda\tau}{f}\right)}{12(1-f)(1-\frac{\lambda\tau}{f})^2}
-1\right]^{1/2}
\end{equation}
At the Pareto-dominant Nash equilibrium, we have
$\tau=\frac{M}{\Omega^*}$ and $\gamma=\gamma^*$. Therefore, based
on \eqref{eq20}, we have
\begin{equation}\label{eq37}
\frac{\lambda\tau}{f}= 2\left[1 + \frac{1}{D\lambda} +
\sqrt{1+\frac{2(1-f^*)}{D\lambda}+
\left(\frac{1}{D\lambda}\right)^2}\right]^{-1} .
\end{equation}
Since $f^*$ is fixed and \eqref{eq37} only depends on the product
$D\lambda$, then $\nu$ and $\chi$ also depend only on the product
of $D$ and $\lambda$, not their individual values. Recall that
$\lambda$ is the average source rate (in packets per second) and
$D$ is the average delay bound. Together, they specify the QoS
requirements of a user. Let $d=D\lambda$. So, for example, if the
packet size $M$ is 100 bits, a source rate of $r=50$kbps results
in $\lambda=500$pps. Then if the delay bound $D$ is 50ms, we have
$d=25$. Fig.~\ref{ratios} shows the plots of $\nu$ and $\chi$
versus $d$ for $f(\gamma)=(1-e^{-\gamma})^M$.

Two important observations can be made from Fig.~\ref{ratios}.
First of all, for moderate and large values of $d$ (e.g., $d>10$),
the average delay is dominated by the average wait time in the
queue (i.e., $\bar{W}^{(q)})$. When $d$ is small, the average wait
time in the queue and the average service time are comparable. For
very small values of $d$, the service time dominates the total
delay. Secondly, for most values of $d$ (i.e., $d>4$), the
standard deviation of ${W}^{(q)}$ is at least ten times larger
than that of $S$. This means that the variations in the total
delay are caused mainly by the variations in $\bar{W}^{(q)}$.
Therefore, in many cases, the variations in the total delay can be
accurately approximated by the variations in the queuing delay.

Now let $w^{(q)}(t)$ be the PDF of the queuing delay. According to
\eqref{eq34c}, the Laplace transform of $w^{(q)}(t)$ is given by
$$W^{(q)*}(s) = \frac{ (1-\rho)s\left(e^{s\tau} -1 +
    f\right)}{s (e^{s\tau} -1 +f) - \lambda
(e^{s\tau} -1)}.$$

We can equivalently write $W^{(q)*}(s)$ as
$$W^{(q)*}(s)=P_0(s)+P_1(s) + P_2(s)$$ where
\begin{equation}\label{eq38}
P_0(s)=(1-\rho) ,
\end{equation}
\begin{equation}\label{eq39}
   P_1(s) =\frac{ (1-\rho) \lambda (e^{s\tau} -1)}{s (e^{s\tau} -1
   +f)} ,
\end{equation}
and
\begin{equation}\label{eq40}
    P_2(s)=\frac{ (1-\rho) \lambda^2 (e^{s\tau} -1)^2}{s
\left[ s (e^{s\tau} -1 + f) -\lambda (e^{s\tau} -1)\right](e^{s\tau}
-1 + f)}.
\end{equation}
Based on \eqref{eq38}, we have
\begin{equation}\label{eq41}
    p_1(t)=(1-\rho) \delta(t) .
\end{equation}

\begin{proposition}\label{prop3}
The inverse Laplace transform of \eqref{eq39} is given by
\begin{equation}\label{eq42}
    p_1(t)= \lambda (1-\rho) (1-f)^{\lfloor\frac{t}{\tau}\rfloor} ,
\end{equation}
where $\lfloor x \rfloor$ represents the nearest integer smaller
than $x$.
\end{proposition}
\begin{thmproof}{Proof:}
See the appendix for the proof.\vspace{0.2cm}
\end{thmproof}
As a result of Proposition~\ref{prop3}, we have
\begin{equation}\label{eq43}
    w^{(q)}(t)= (1-\rho)\delta(t) + \lambda (1-\rho)
    (1-f)^{\lfloor\frac{t}{\tau}\rfloor} + p_2(t) .
\end{equation}
Now if we restrict our attention to $0\leq t \leq t_{max}$ where
$t_{max}>> D$, then we can approximate $p_2(t)$ numerically using
the following:
$$P_2(i\omega)= \int_0^{t_{max}} p_2(t) e^{-i\omega t} \textrm{d}t \simeq \sum_{n=0}^{N-1} p_2\left(\frac{t_{max}}{N}n\right)
e^{-i\omega \frac{t_{max}}{N}n} \left(\frac{t_{max}}{N}\right)$$ or
$$\left(\frac{N}{t_{max}}\right)P_2(i\omega) =\sum_{n=0}^{N-1} p_2\left(\frac{t_{max}}{N}n\right)
e^{-i\omega \frac{t_{max}}{N}n} .$$

Now, since the FFT of a discrete signal $z_n$ is given by
$$Z_k = \sum_{n=0}^{N-1} z_n
e^{-i\frac{2\pi k n}{N}} ,$$ $p_2\left(\frac{t_{max}}{N}n\right)$
can be obtained by taking the IFFT of
$\left(\frac{N}{t_{max}}\right) P_2(s)|_{s=i\frac{2\pi
k}{100D}}$\footnote{Since $p_2(t)$ is real, before taking the IFFT,
we have to make sure that the samples of $P_2(s)$ satisfy the
symmetry properties associated with the FFT of real signals.}. In
Section~\ref{numerical results}, we use this approximation along
with \eqref{eq43} to obtain $w^{(q)}(t)$ and, consequently,
approximate $\text{Pr}\{W^{(q)}\leq c\}$. This allows us to quantify
the delay performance of the users at Nash equilibrium.

\section{Numerical Results}\label{numerical results}

Let us consider the uplink of a DS-CDMA system with a total
bandwidth of 5MHz (i.e. $B=5$MHz). A useful example for the
efficiency function is ${f(\gamma)= (1- e^{-\gamma})^M}$. This
serves as an approximation to the packet success rate that is very
reasonable for moderate to large values of $M$. We use this
efficiency function for our simulations. Using this, with $M=100$,
we have $\gamma^*=6.48 = 8.1$dB. Each user in the network has a
set of QoS requirements expressed as $(r_k, D_k)$ where $r_k$ is
the source rate and $D_k$ is the delay requirement (upper bound on
the average total delay) for user~$k$. As explained in
Section~\ref{NE}, the QoS parameters of a user define a ``size"
for that user, denoted by $\Phi_k^*$ given by \eqref{eq24}. Before
a user starts transmitting, it must announce its size to the
access point. Based on the particular admission policy, the access
point decides whether or not to admit the user. Throughout this
section, we assume that the admitted users choose the transmit
powers and rates that correspond to their Pareto-dominant Nash
equilibrium.

\begin{figure}
\centering
\includegraphics[width=5in]{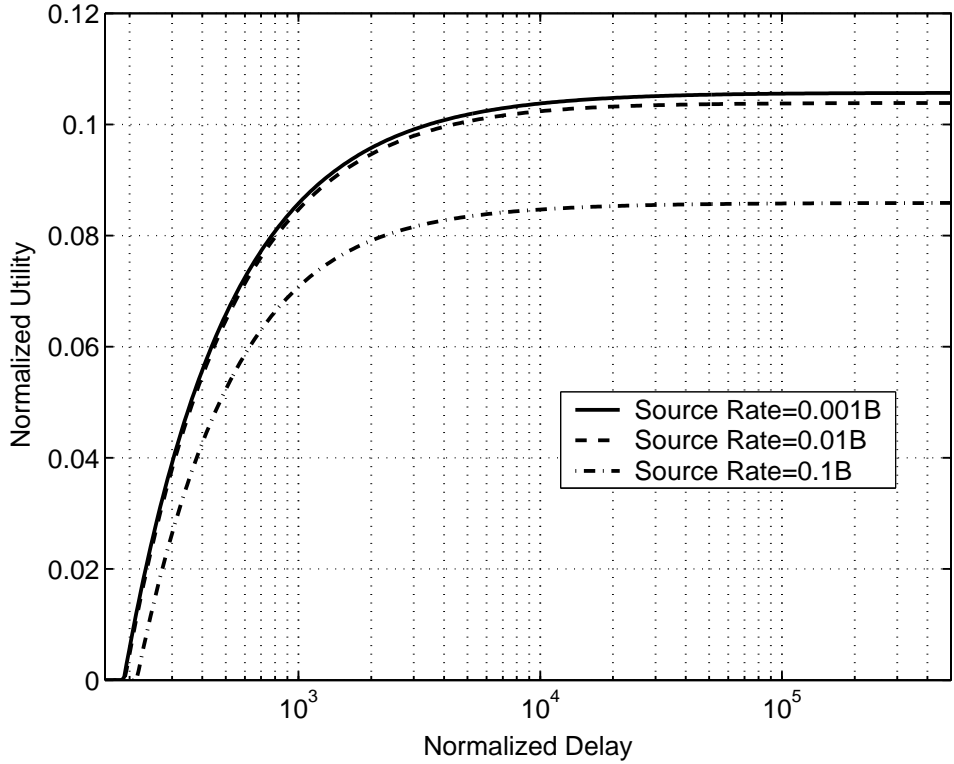}
\caption{Normalized utility as a function of normalized delay for
different source rates ($B=5$~MHz). The combined ``size" of other
users in the network is equal to 0.2.} \label{utilvsdelay}
\end{figure}

Fig.~\ref{utilvsdelay} shows the user's utility as a function of
delay for different source rates. The total size of the other
users in the network is assumed to be 0.2. The user's utility is
normalized by $Bh/\sigma^2$, and the delay is normalized by the
inverse of the system bandwidth. As expected, a tighter delay
requirement and/or a higher source rate results in a lower utility
for the user.

\begin{figure}
\centering
\includegraphics[width=5in]{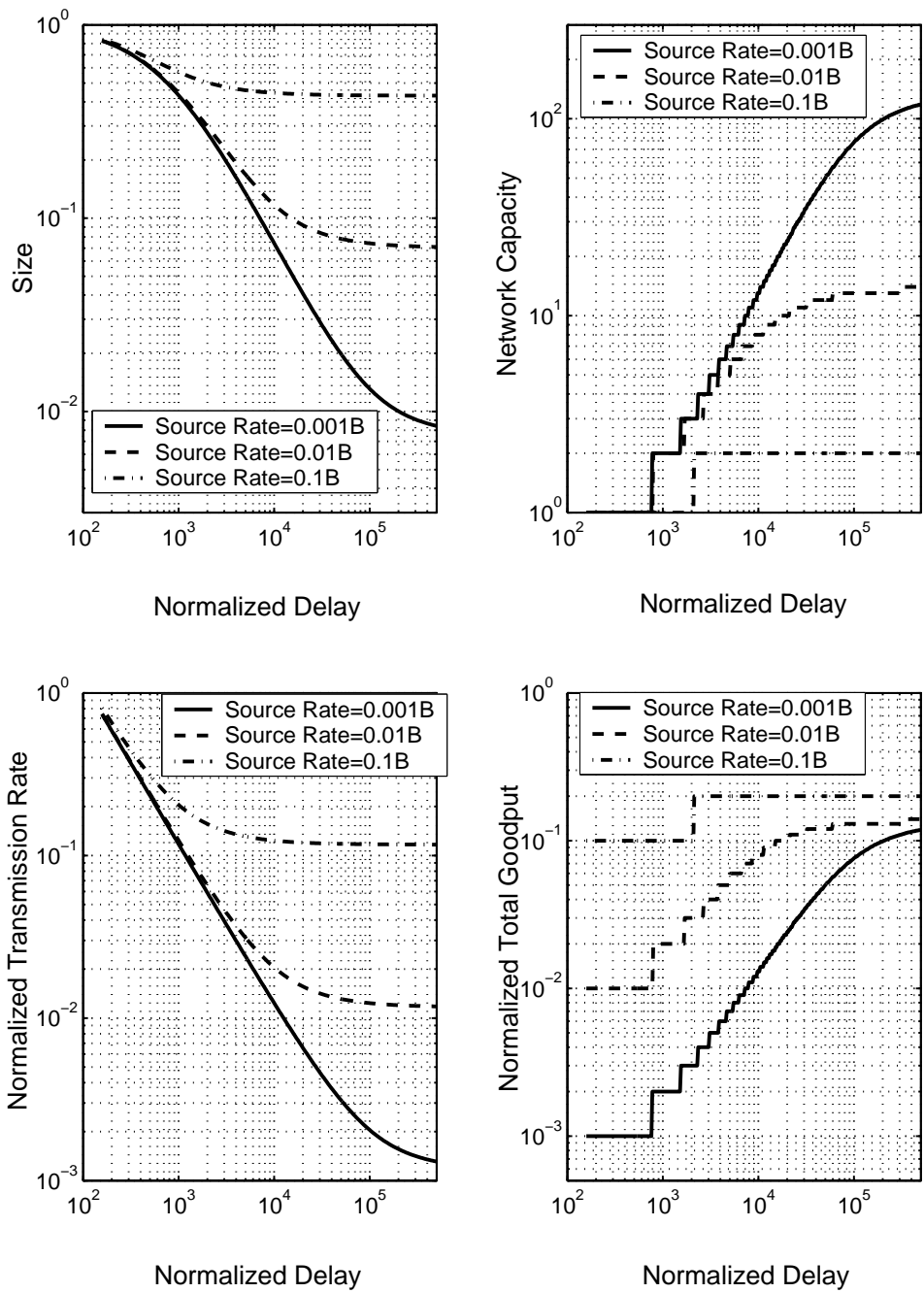}
\caption{User size, network capacity, normalized transmission
rate, and normalized total goodput as a function of normalized
delay for different source rates ($B=5$~MHz).} \label{delayQoS}
\end{figure}

Fig.~\ref{delayQoS} shows the user size, network capacity,
transmission rate, and total goodput as a function of normalized
delay for different source rates. The network capacity refers to
the maximum number of users that can be admitted into the network
assuming that all the users have the same QoS requirements (i.e.,
the same size). The transmission rate and goodput are normalized
by the system bandwidth. The total goodput is obtained by
multiplying the source rate by the total number of users. For
example, a user with a source rate of 50~kbps and an average delay
constraint of 50~ms (i.e., $r=50$~kbps and $D=50$~ms) has a size
equal to 0.072. As the QoS requirements become more stringent
(i.e., a higher source rate and/or a smaller delay), the size of
the user increases which means more network resources are required
to accommodate the user. This results in a reduction in the
network capacity. For $r=50$~kbps and $D=50$~ms, the transmission
rate is equal to 59.65~kbps, the network capacity is equal to 13,
and the total goodput is 650~kbps. It is also observed from the
figure that when the delay constraint is loose, the total goodput
is almost independent of the source rate. This is because a lower
source rate is compensated by the fact that more users can be
admitted into the network. On the other hand, when the delay
constraint in tight, the total goodput is higher for larger source
rates.

Now, to study admission control, let us consider a network with
three different classes of users/sources:
\begin{enumerate}
  \item Class $A$ users for which $r^{(A)}=5$~kbps and $D^{(A)}=10$~ms.
  \item Class $B$ users for which $r^{(B)}=50$~kbps and $D^{(B)}=50$~ms.
  \item Class $C$ users for which $r^{(C)}=150$~kbps and $D^{(C)}=1000$~ms.
\end{enumerate}
We can calculate the size of a user in each class using
\eqref{eq24} to get $\Phi^{*^{(A)}}=0.0198$,
$\Phi^{*^{(B)}}=0.0718$, and $\Phi^{*^{(C)}}=0.1848$. This means
that users in classes $B$ and $C$ respectively consume
approximately 3.6 and 9.3 times as much resources as a user in
class $A$.

For the purpose of illustration and to keep the comparison fair,
let us assume that there are a large number of users in each class
and that they all are at the same distance from the access point
(i.e., they all have the same average channel gain). The access
point receives requests from the users and has to decide which
ones to admit in order to maximize the total utility in the
network (see \eqref{eq29}). We know from Section~\ref{admission
control} that since users in class $A$ have the smallest size, the
total utility is maximized if the access point picks users from
class $A$ only with $L^{(A)}= \left[1/2\Phi^{*^(A)}\right]=25$.
However, this solution does not take into account fairness.
Instead, we may be more interested in cases where more than one
class of users are admitted. Table~\ref{table1} shows the
percentage loss in the total utility (energy efficiency) for
several choices of $L^{(A)}, L^{(B)}$ and $L^{(C)}$. It is
observed that admitting ``large" users into the network results in
significant reductions in the energy efficiency and capacity of
the network.
\begin{table}[t]
\begin{center} \caption{ Percentage loss in the total network utility for different choices of $L^{(A)},
L^{(B)}$ and $L^{(C)}.$}\label{table1}
\begin{tabular}{|c|c|c|c|}
  \hline
  $L^{(A)}$ & $L^{(B)}$ & $L^{(C)}$ & Loss in total utility \\
  \hline
  \hline
  25 & 0 & 0 & -- \\
   23 & 1 & 0 & 10\% \\
  20 & 0 & 1 & 30\% \\
  18 & 1 & 1 &  38\% \\
  0 & 7 & 0 & 71\% \\
  0 & 0 & 3 & 87\% \\
  \hline
\end{tabular}
\end{center}
\end{table}

\begin{figure}
\centering
\includegraphics[width=5in]{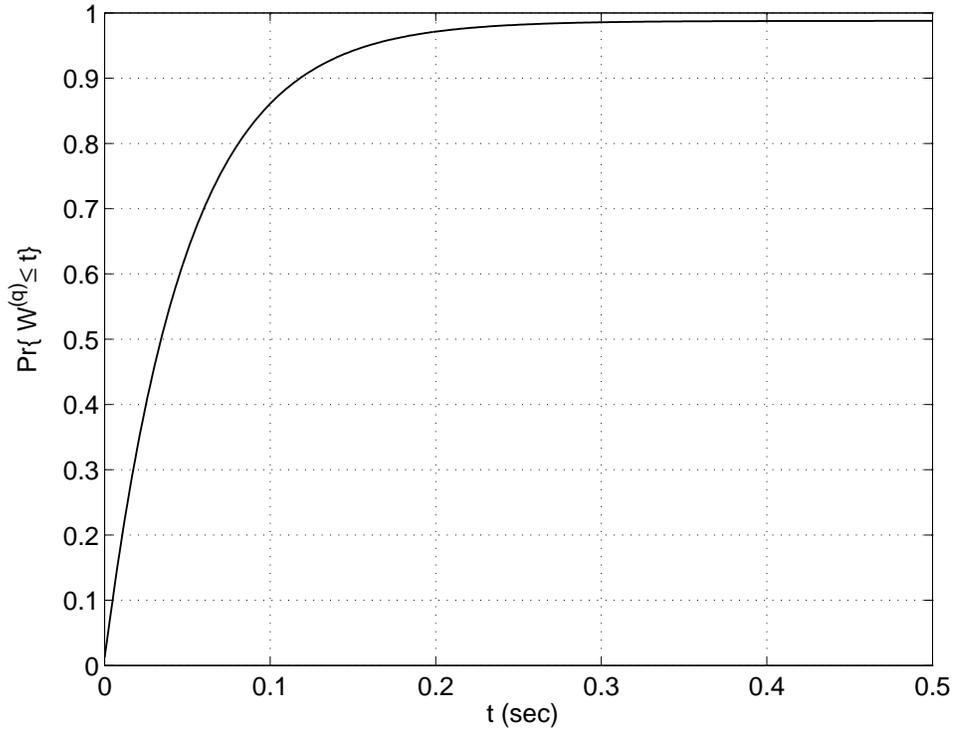} \caption{Cumulative distribution function of the
queuing delay for a user with a source rate of 50~kbps and an
average delay of 50~ms.} \label{cdfdelay}
\end{figure}

Let us now focus on the delay profile of a user in class $B$. For
this user, we have $r^{(B)}=50$~kbps (or $\lambda^{(B)}=500$~pps)
and $D^{(B)}=50$~ms. Therefore, $d^{(B)}=25$. From
\eqref{eq34d}--\eqref{eq34g}, we have $\bar{S}^{(B)}=2$~ms,
$\sigma_S^{(B)} =0.74$~ms, $\bar{W}^{(q)(B)}=48$~ms and
$\sigma_{W^{(q)}}^{(B)}=48$~ms. It is clear that for this user the
queuing delay is the dominant component of the total delay. This
can also be seen from Fig.~\ref{ratios}. Therefore, the cumulative
distribution function (CDF) of $W^{(B)}$, i.e.,
$\text{Pr}\{W^{(B)}\leq t\}$, can be very accurately approximated
by the CDF of $W^{(q)(B)}$. Hence, we can use \eqref{eq43} to
numerically compute the CDF of the queuing delay. This CDF is
plotted in Fig.~\ref{cdfdelay}. It is seen from the figure that
about 63\% of the time, the delay experienced by a packet is less
than the average delay bound and 85\% of the time, the delay is
less than twice the average delay.

\section{Conclusions}\label{conclusions}

We have studied the cross-layer problem of QoS-constrained power
and rate control in wireless networks using a game-theoretic
framework. We have proposed a non-cooperative game in which users
seek to choose their transmit powers and rates in such a way as to
maximize their utilities and at the same time satisfy their QoS
requirements. The utility function considered here measures the
number of reliable bits transmitted per joule of energy consumed.
The QoS requirements for a user consist of the average source rate
and an upper bound on the average delay where the delay includes
both transmission and queuing delays. We have derived the Nash
equilibrium solution for the proposed game and obtained a
closed-form solution for the user's utility at equilibrium. Using
this framework, we have studied the tradeoffs among throughput,
delay, network capacity and energy efficiency, and have shown that
the presence of users with stringent QoS requirements results in
significant reductions in network capacity and energy efficiency.
The delay performance of users at Nash equilibrium are also
analyzed.

\appendix

\section*{Proof of Proposition~\ref{prop3}}

Given $P_1(s)=\frac{ (1-\rho) \lambda (e^{s\tau} -1)}{s (e^{s\tau}
-1 +f)}$, we can use inverse Laplace transform to write
$$p_1(t)= \lim_{R\rightarrow \infty} \frac{1}{2\pi i}
\int_{\sigma-iR}^{\sigma+iR} P_1(s) e^{st} ds .$$

Using the residue theorem and contour integration from complex
analysis \cite{WunschBook94}, we have
$$p_1(t)= \sum_k \textrm{Res}\left[P_1(s) e^{st} , s_k^*\right]$$ where
$s_k^*= \frac{1}{\tau} \left[\ln(1-f) + 2\pi i k\right]$.

If we let $a = \ln (1-f)$, then we have
$$p_1(t)= (1-\rho)\lambda \sum_{k=-\infty}^{\infty} \frac{(e^a-1)
e^{at/\tau + 2\pi i k t/\tau}}{(1-f) (a+2\pi i k)}.$$

For convenience, let us define $x= t/\tau$ and notice that $x\geq0$
since the queuing delay is non-negative. Then, we can write
$$p_1(t)=-f (1-\rho)\lambda (1-f)^{x-1} \sum_{k=-\infty}^{\infty} \frac{
e^{2\pi i k x}}{ a+2\pi i k}= -f (1-\rho)\lambda (1-f)^{x-1}
\sum_{k=-\infty}^{\infty} \frac{(a-2\pi ik) e^{2\pi i k x}}{
a^2+4\pi^2 k^2}.$$

Define $h(x)=\sum_{k=-\infty}^{\infty} \frac{(a-2\pi ik) e^{2\pi i k
x}}{ a^2+4\pi^2 k^2}$. Then, we have
\begin{equation}\label{eqApp1a}
 p_1(t)=-f (1-\rho)\lambda (1-f)^{x-1} h(x) .
\end{equation}
We can rewrite $h(x)$ as
$$h(x) = \frac{1}{2\pi} \left[ \sum_{k=-\infty}^{\infty} \frac{b e^{2\pi i k x}}{
b^2+ k^2} - i \sum_{k=-\infty}^{\infty} \frac{k e^{2\pi i k x}}{
b^2+ k^2}\right]$$ where $b=\frac{a}{2\pi}$. We can equivalently
write $h(x)$ as
\begin{equation}\label{eqApp1}
h(x) = \frac{1}{2\pi b}+ \frac{1}{\pi} \left[ \sum_{k=1}^{\infty}
\frac{ b\cos{(2\pi k x)}}{ b^2+ k^2} + \sum_{k=1}^{\infty} \frac{k
\sin{(2\pi  k x)}}{ b^2+ k^2}\right] .
\end{equation}

Now, given the following Fourier series expansions
\cite{TableBook63}
$$\sum_{k=1}^{\infty}
\frac{ \cos{(k y)}}{ b^2+ k^2}=
\frac{\pi}{2b}\frac{e^{b(\pi-y)}+e^{-b(\pi-y)}}{e^{b\pi}-e^{-b\pi}}
-\frac{1}{2b^2} \ \ \ \textrm{for} \ \ \ 0<y<2\pi$$ and
$$\sum_{k=1}^{\infty}
\frac{ k\sin{(k y)}}{ b^2+ k^2}=
\frac{\pi}{2}\frac{e^{b(\pi-y)}-e^{-b(\pi-y)}}{e^{b\pi}-e^{-b\pi}}
\ \ \ \ \ \ \ \textrm{for} \ \ \ 0<y<2\pi$$ and after some
manipulations, $h(x)$ becomes

$$h(x) =\frac{e^{-2\pi b(x-n)}}{1-e^{-2\pi b}}  \ \ \ \ \textrm{for} \ \ n<x<n+1
.$$ Remembering that $a=\ln(1-f)$, we can simplify $h(x)$ to get
\begin{equation}\label{eqApp2}
    h(x)=\frac{(1-f)(1-f)^{-(x-n)}}{-f}  \ \ \ \ \textrm{for} \ \ n<x<n+1 .
\end{equation}
Since $p_1(t) =-f (1-\rho)\lambda (1-f)^{x-1} h(x)$ and recalling
that $x=\frac{t}{\tau}$, we get
$$p_1(t)=\lambda(1-\rho) (1-f)^n   \ \ \ \ \textrm{for} \ \
n\tau<t<(n+1)\tau$$ or equivalently\vspace{0cm}
$$p_1(t)=\lambda(1-\rho) (1-f)^{\lfloor\frac{t}{\tau}\rfloor}
.\vspace{0cm}$$ This completes the proof. 

\end{document}